\documentstyle[twocolumn,prb,aps,epsf]{revtex}
\begin{document}
\draft
\def \beq{\begin{equation}}
\def \eeq{\end{equation}}
\def \beqarr{\begin{eqnarray}}
\def \eeqarr{\end{eqnarray}}

\twocolumn[\hsize\textwidth\columnwidth\hsize\csname @twocolumnfalse\endcsname
\title{Zeeman and Orbital Effects of an in-Plane Magnetic Field in Cuprate 
Superconductors}

\author{Kun Yang}

\address{
%Condensed Matter Physics 114-36, California Institute of Technology,
%Pasadena, California 91125\\
National High Magnetic Field Laboratory and Department of Physics,
Florida State University, Tallahassee, Florida 32310
}

\author{S. L. Sondhi}
\address{
Department of Physics,
Princeton University,
Princeton, New Jersey 08544
}

\date{\today}
\maketitle
\begin{abstract}

We discuss the effects of a magnetic field 
applied parallel to the Cu-O ($ab$) 
plane of the high $T_c$ cuprate superconductors.
After briefly reviewing the Zeeman effect of the field, we study the 
orbital effects, using the Lawrence-Doniach model for
layered superconductors as a guide to the
physics. We argue that the orbital effect is qualitatively 
different for in-plane and inter-layer mechanisms for superconductivity.
In the case of in-plane mechanisms, interlayer couplings may be modeled
as a weak interlayer Josephson coupling, whose effects disappear as 
$H\rightarrow\infty$; in this case Zeeman dominates the effect of the field.
In contrast, in the inter-layer mechanism  
the Josephson coupling {\em is} the driving force of superconductivity,
and we argue that
the in-plane field suppresses superconductivity and provides an upper
bound for $H_{c2}$ which we estimate very crudely.

\end{abstract}

]

One of the most important milestones in the study of the cuprate superconductors
is the identification of the predominantly $d_{x^2-y^2}$ symmetry of their
pairing order parameters.
This discovery has motivated a tremendous amount of
theoretical work on the
physical properties of superconductors with 
unconventional pairing symmetry.

Among them, it was recently pointed out\cite{yang} that
the response of a two-dimensional (2D)
$d_{x^2-y^2}$ superconductor to the Zeeman coupling
between the spins of the electrons and an external  
magnetic field is  much stronger than that in an 
$s$-wave superconductor. This is so, for unlike the fully gapped
$s$-wave case, in a $d_{x^2-y^2}$ superconductor 
there exists nodal points in 
the order parameter and therefore gapless quasiparticle excitations. Consequently,
an arbitrarily weak Zeeman field generates a finite density of spin-polarized 
quasiparticles in the ground state\cite{yang}. 
%These 
%quasiparticles form Fermi seas of their own near the nodal points, and 
%dominate the low-temperature thermodynamic properties of the system. 
It was also
shown that the Zeeman field enhances the low-$T$ specific heat, 
thermal conductivity, and tunneling density of states, 
while suppressing the superfluid density\cite{yang,won}.
At strong fields, the system supports a Fulde-Ferrell-Larkin-Ovchinnikov
type superconducting state.\cite{maki,yang}  
Eventually, a sufficiently strong Zeeman field suppresses
superconductivity completely, leading to a ``Pauli limit" on the upper critical
field.

The results mentioned above are expected to be particularly relevant to the cuprate 
superconductors  when the external field is oriented {\em parallel} to the Cu-O 
($ab$) planes (for possible caveats on the influence of disorder see
[\onlinecite{grimaldi}]), as the cuprates are quasi-2D systems with electrons 
predominantly moving in the Cu-O planes. In a {\em truly} 2D system,
the orbital motions of the electrons would not
sense the existence of a parallel magnetic field, and the Zeeman effect 
would be the {\it only} effect of the field.
Experimentally, there is some evidence\cite{obrein}
suggesting that the upper critical field for the parallel orientation 
is indeed Pauli limited in YBCO at low temperatures.

However for smaller fields  there do exist
orbital effects, even when the field is parallel to the planes, due to the
presence of inter-plane couplings.
It is the purpose of this paper to report results of studies of such 
orbital effects, and to compare them with the Zeeman effect in field ranges
that are relevant to current experiments. We will argue that the nature
and importance of the orbital effects depend sensitively on whether 
superconductivity is driven by in-plane correlations or inter-plane hopping. 
Thus their study may provide a way to  distinguish between in-plane and inter-plane
mechanisms for superconductivity in the cuprates.

The orbital effects of the parallel field may be summarized qualitatively as 
the following. The field generates screening currents both within and 
across the planes. The current within the plane generates a ``Doppler shift"
of the quasiparticle spectrum\cite{yip}, and therefore a finite 
density of states (DOS) for the quasiparticles. This current-induced DOS is 
proportional to $\sqrt{H}$ for $H\Vert \hat{c}$\cite{volovik}, and thus 
dominates the Zeeman effect (which gives rise to a quasiparticle DOS that
is proportional to $H$\cite{yang}) at low $H$. We show below that the situation is
quite different for an in-plane field ($H\bot\hat{c}$). 
Also the inter-plane Josephson current suppresses the inter-plane Josephson
coupling energy. Thus, depending on how crucial the Josephson coupling energy
is to superconductivity itself, the response of the system to the in-plane
field can be very different.

We model the cuprates by the Lawrence-Doniach model\cite{ld} that is 
appropriate for layered superconductors\cite{note1}:

\beqarr
&F&=\sum_n\int{dxdy}\{\alpha|\psi_n|^2+{\beta\over 2}|\psi_n|^4
\nonumber\\
&+&{\hbar^2\over 2m_{ab}}|(\nabla-{2ie\over \hbar c}{\bf A}_n)\psi_n|^2
+t|\psi_{n+1}-\psi_n|^2\}.
\label{ld}
\eeqarr
Here $n$ is the layer index, $\alpha\propto T-T_c$, and other parameters may 
be assumed to be $T$ independent. For $T$ very close to $T_c$, 
the coherence length becomes very large, so that continuum approximation
can be used along the $c$ direction, and (\ref{ld}) reduces to the usual 
Ginsburg-Landau theory with anisotropic mass, and the mass along the $c$
direction is $m_c={\hbar^2\over 2s^2t}$ where $s$ is the spacing between 
layers. The parameter $\gamma^2=m_c/m_{ab}$ is a measure of the degree
of anisotropy of the system; for examples, $\gamma^2\approx 50$ for YBCO, and
$\gamma^2\approx 20000$ for BSCCO\cite{tinkham}.

We are primarily interested in the low $T$ regime, where the layered structure
is important; more precisely we are interested in the case $T < T^*$, the
latter being the temperature at which the orbital $H_{c2}$ diverges. 
At such temperatures we distinguish two regimes: a low field regime in which 
the current flowing in the layers increases with $H$ on account of increasing vortex density, and a high field regime in which the increasing density is 
counteracted by the cancelation between neighboring vortex patterns.
The former has also been discussed by Volovik\cite{volovik} and also yields 
a DOS $\propto\sqrt{H}$. Here we compute the answer at high fields.
We assume the external field ${\bf H}=H(1, 0, 0)$ is along the
$\hat{a}$ direction. It is known that $H_{c1}$  is very low 
($ < 100 Gauss$) for a parallel field, and for fields of order $H\sim 1T$
(which is the range of interest in the present paper and typical for
experimental studies), the field basically 
penetrates the system uniformly and the screening effect is negligible. 
(This justifies neglecting the 
field energy term in (\ref{ld})). Thus we can assume the field inside the
superconductor is the same as that of the external field and uniform. We 
use the gauge ${\bf A}_n=nsH(0, 1, 0)$.

We need to solve the equation for $\psi_n$:
\beqarr
{\delta F\over \delta \psi_n^*}&=&\alpha\psi_n+\beta|\psi_n|^2\psi_n
-{\hbar^2\over 2m_{ab}}(\nabla-{2ie\over \hbar c}{\bf A}_n)^2\psi_n
\nonumber\\
&-&t(\psi_{n-1}+\psi_{n+1})=0.
\label{sch}
\eeqarr
For in-plane mechanisms, the Josephson coupling $t$ is weak and secondary to
superconductivity. To find the solution to (\ref{ld}), we first set $t=0$, and
then treat $t$ as a weak perturbation. For $t=0$, all the layers decouple, 
and we obtain
\beq
\psi_n(x, y)=\sqrt{-\alpha/\beta}e^{i(\phi_n+k_ny)},
\eeq 
where $k_n={2ensH\over\hbar c}$ and
$\phi_n$ are arbitrary phases. Upon introducing $t$ these solutions spread to
more than one layer {\em and} lock phases, leading to the unique,
leading order solution\cite{note2}
\beq
\psi_n(x, y)=\sqrt{-\alpha\over \beta}
e^{ik_ny}[\eta_n+{2i\eta_{n+1}\over \gamma^2}
({\ell\over s})^4\sin(\Delta ky)],
\eeq
where $\Delta k=2esH/\hbar c$, $\ell=\sqrt{\hbar c/2eH}$ is the magnetic
length for Cooper pairs, 
$\eta_n=1$ for $n=4j$ or $4j+1$, and $-1$ otherwise ($j$ is an integer).
In order for the perturbative approach to be valid, we need to have
\beq
{1\over \gamma^2}
({\ell\over s})^4 \ll 1,
\label{condition}
\eeq
which is always valid in the strong field limit.

The typical in-plane superfluid velocity is thus
\beq
v_s\approx {\hbar\Delta k\over \gamma^2 m_{ab}}({\ell\over s})^4,
\eeq
and the Doppler shift energy is\cite{yip,volovik}
\beq
E_D\sim\hbar v_sk_F={\epsilon_F\over \gamma^2}{\ell^2\over k_Fs^3}\propto 1/H.
\eeq 
Since the Zeeman energy $E_Z=\mu_BH\propto H$, it always dominates the 
Doppler shift at sufficiently strong field. At field $H=10 T$, using
$\epsilon_F\sim 2 eV$, $k_F\sim 1{\rm \AA}^{-1}$ and $s\sim 10 {\rm \AA}$, 
we obtain 
for BSCCO (in which (\ref{condition}) is satisfied)
$E_D/E_Z\sim 0.4$,
indicating Zeeman effect starts to dominate the orbital effect in this field
range in BSCCO. (\ref{condition}) is {\em not} satisfied in YBCO at this 
field due to less anisotropy; 
but using the above results anyway (which results in an {\em overestimate} of
the in-plane current) we find $E_D/E_Z \sim 150$, 
suggesting that the orbital effect will dominate.

In the above discussion we have {\em assumed} that the interlayer Josephson
couplings are secondary to superconductivity in cuprates, and may be treated
as weak perturbations. This is appropriate for in-plane mechanisms for
superconductivity in cuprates. However it was proposed by Anderson and
coworkers\cite{anderson,chakravarty} that the interlayer Josephson coupling
(or pair-hopping) may actually be the driving force behind superconductivity
in cuprates. In this case, the interlayer Josephson coupling can no longer
be treated as a weak perturbation as above, and we expect an in-plane 
magnetic field to have a much more dramatic orbital effect.
This is because the in-plane field tends to suppress interlayer phase 
coherence,
and therefore Josephson energy and superconductivity in this case. Among other
things, the orbital effect of the field provides a mechanism  for the upper
critical field $H_{c2}$. While for in-plane mechanisms, the field 
simply decouples the layers, and $H_{c2}$ can only come from the 
Zeeman effect\cite{yang,maki}.

In the inter-layer pair hopping model\cite{anderson,chakravarty}, the 
multi-layer structure of the cuprates is important. In particular, 
the Josephson coupling among layers within the same unit cell might be
much stronger than between different cells, thus the large anisotropy
could largely reflect the weakness of the inter-unit cell Josephson coupling
only.
For simplicity, we study in the following in a single bilayer (as in Ref.
\onlinecite{chakravarty}), and neglect the coupling between this pair of layers 
and the rest of the system. 
A proper treatment of even this problem requires a microscopic theory 
that does not really exist. 
So here we will content ourselves with a speculative
illustration of what such a treatment might produce. To this end
we again use the Lawrence-Doniach model, in a slightly
different form and with a crucial {\em reinterpretation} of the parameters:
\beqarr
F&=&\int{dxdy}\{\alpha'(|\psi_1|^2+|\psi_2|^2)+{\beta\over 2}
(|\psi_1|^4+|\psi_2|^4)
\nonumber\\
&+&{\hbar^2\over 2m_{ab}}(|\nabla \psi_1|^2|+|\nabla \psi_2|^2|)
\nonumber\\
&-&t(\psi_1^*\psi_2e^{{2ieHsx\over \hbar c}}+\psi_2^*\psi_1e^{-{2ieHsx
\over \hbar c}})\}.
\label{ldd}
\eeqarr
Here $\alpha'=\alpha + t$. We have used a different gauge: For a parallel
field ${\bf H}=H\hat{y}$, we choose ${\bf A}=Hx\hat{z}$, so the ${\bf A}$
affects the phase of the interlayer Josephson coupling, but not in-plane
kinetic energy. The physics, of course, is unaffected by the choice of gauge.
The reason to isolate the last term out is that this term 
describes interlayer coupling {\em only}; while all the
in-plane properties are described by other terms.

For in-plane mechanisms, each individual layer is superconducting (at low $T$)
without interlayer coupling. Thus we must have $\alpha' < 0$, which means 
the reason that a superconducting order parameter $\psi$ is developed is 
because of the lowering of in-plane free energy. Without a parallel field, 
the free energy can be further lowered by having the {\em same} phase for
$\psi_1$ and $\psi_2$ (and uniform throughout the system so that there is no
in-plane kinetic energy cost), so that the systems gains Josephson coupling
energy as well.

In pair hopping mechanism, however, the individual layers are {\em not}
superconducting without interlayer Josephson coupling. This means
$\alpha' > 0$! Instead the system becomes superconducting at $H=0$
by developing a non-zero $\psi$ in {\em both} layers, {\em and} making them 
phase coherent ($\psi_1=\psi_2$), thereby gaining  
Josephson coupling energy. Thus as long as $t > \alpha'$,
the system is in the superconducting phase, despite $\alpha' > 0$.

As long as $H=0$, the ground state of (\ref{ldd}) is the same:
$\psi_1=\psi_2=\sqrt{(t-\alpha')/\beta}$, and there is no obvious distinction
between these two cases; 
the system is phase coherent both within each individual
layer and between the two layers.
The situation is very different when $H\ne 0$ and is sufficiently large. This
is because for a loop that encloses a finite amount of flux, one must accumulate
a finite amount of gauge-invariant phase difference proportional to the flux
(see Fig. 1a).
It is therefore no-longer possible to maintain in-plane and interlayer phase
coherence simultaneously\cite{yang2}.

For $\alpha' < 0$ (in-plane mechanism), the situation is basically the same as
a Josephson junction in a magnetic field\cite{tinkham}; we
have $\psi_j=e^{i\phi_j}\sqrt{-\alpha'/\beta}$, with no correlation between
$\phi_1$ and $\phi_2$. The leading order correction to this is proportional to
$t/H$. The field induces an oscillatory Josephson current between the layers:
$j_z\approx (-\alpha't/\beta)\sin({2esHx\over \hbar c}+\phi_1-\phi_2)$, 
while the in-plane current goes to zero as $1/H$ at large $H$ (see Fig. 1b). 
The Josephson coupling term in (\ref{ldd}) averages to zero in this solution. 
In short, the system gives up interlayer coherence (as manifested by the
loss of Josephson coupling energy and appearance of Josephson current), in 
order to maintain in-plane coherence (therefore no in-plane current and
kinetic energy).

For $\alpha' > 0$ (interlayer mechanism), on the other hand, the system will do
everything possible to maintain interlayer coherence, as the Josephson
energy is the origin of superconductivity. Thus the solution in this case is
\beq
\psi_{1,2}=\sqrt{(t+\alpha'-{(esH)^2\over 2m_{ab}c^2})/\beta
}
e^{\pm iesHx/\hbar c}. 
\label{sol}
\eeq
In this solution the interlayer coherence is maintained
and there in {\em no} interlayer Josephson current induced by the field. 
As a price, the system loses in-plane coherence, and there is diamagnetic
current ($\propto H$) and in-plane kinetic energy loss ($\propto H^2$)
(see Fig. 1c).
Clearly, as $H$ increases, at some point the loss of in-plane kinetic energy
will overwhelm the gain of Josephson energy, and the system will cease to
superconduct. This is the orbital upper critical field $H_{c2}^o$
(to be distinguished from the Zeeman upper critical field $H_{c2}^Z$), as determined
by the point where $\psi$ vanishes in (\ref{sol}).
For in-plane mechanisms, $H_{c2}^o\rightarrow\infty$. If we {\em assume} that
interlayer mechanism is responsible for superconductivity in cuprates, 
a crude estimate yields in BSCCO2212 $H_{c2}^o\sim 200 T$ at $T=0$. 
Unfortunately this is comparable to $H_{c2}^Z$\cite{yang,maki}. However,
in principle, the Zeeman effect can be suppressed by introducing spin-orbital 
scatterers to the system which could enable the two to be distinguished. 

The qualitative difference in the orbital response to a parallel
magnetic field can provide a way to experimentally
distinguish between in-plane and interlayer mechanisms for cuprate 
superconductivity.
We note the interpretation of previous experimental attempts
to distinguish between the two based on $c$-axis penetration depth is
controversial\cite{sudip}.

This work was supported by NSF DMR-9971541 and the Sloan Foundation
(KY), and NSF DMR-9632690, and Sloan and Packard Foundations (SLS).

\vskip -2cm
\begin{figure*}[h]
\centerline{\epsfxsize=12cm
\epsfbox{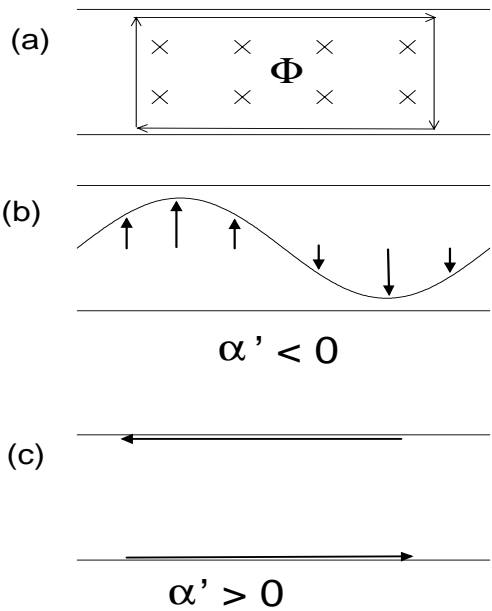}
}
\vskip -2cm
\caption{
(a) A closed path that encloses a finite amount of flux.
(b) The current pattern in a bilayer system in the presence of strong parallel
magnetic field, for in-plane mechanism of superconductivity.
(c) Same as (b), for inter-layer pair hopping
mechanism of superconductivity.
}
\label{fig1}
\end{figure*}


\begin{references}
\bibitem{note1} In the following model it is assumed that each unit cell
along the $\hat{c}$ direction contains one layer, and all layers are 
equivalent. Many cuprates, however, have multiple layers in a unit cell. Such
multi-layered structure is not very important for in-plane mechanisms
for superconductivity, but is very important for the interlayer pair
hopping mechanism\cite{anderson}. See discussions later.
\end{references}
\end{document}